\documentclass[english,reprint,prl]{revtex4-1}
\usepackage{lmodern}

\usepackage[T1]{fontenc}
\usepackage[latin9]{inputenc}
\usepackage{color}
\usepackage{textcomp}
\usepackage{amsmath}
\usepackage{amssymb}
\usepackage{graphicx}
\usepackage{babel}
\begin{document}

\title{Inferring network connectivity from event timing patterns}

\author{Jose Casadiego$^{1,2}$, Dimitra Maoutsa$^{2}$, Marc Timme$^{1,2,3,4}$}

\affiliation{{\footnotesize{}$^{1}$Chair for Network Dynamics, Institute of Theoretical
Physics and Center for Advancing Electronics Dresden (cfaed), Technical
University of Dresden, 01062 Dresden, Germany}~\\
{\footnotesize{}$^{2}$Network Dynamics, Max Planck Institute for
Dynamics and Self-Organization (MPIDS), 37077 Göttingen, Germany }~\\
{\footnotesize{}$^{3}$Bernstein Center for Computational Neuroscience
(BCCN), 37077 Göttingen, Germany}~\\
{\footnotesize{}$^{4}$Advanced Study Group, Max Planck Institute
for the Physics of Complex Systems, 01069 Dresden, Germany}}
\begin{abstract}
Reconstructing network connectivity from the collective dynamics of
a system typically requires access to its complete continuous-time
evolution although these are often experimentally inaccessible. Here
we propose a theory for revealing physical connectivity of networked
systems only from the event time series their intrinsic collective
dynamics generate. Representing the patterns of event timings in an
\emph{event space} spanned by inter-event and cross-event intervals,
we reveal which other units directly influence the inter-event times
of any given unit. For illustration, we linearize an event space mapping
constructed from the spiking patterns in model neural circuits to
reveal the presence or absence of synapses between any pair of neurons
as well as whether the coupling acts in an inhibiting or activating
(excitatory) manner. The proposed model-independent reconstruction
theory is scalable to larger networks and may thus play an important
role in the reconstruction of networks from biology to social science
and engineering.
\end{abstract}
\maketitle
The topology of interactions among the units of network dynamical
systems fundamentally underlies their systemic function. Current approaches
to reveal interaction patterns of a network from the collective nonlinear
dynamics it generates \cite{Yeung:2002,Gardner2003,Yu2006,Timme2007,Yu:2010a,Wang2016,Barzel:2013,Timme2014,Nitzan2017}
rely on directly sampling the trajectories of the collective time
evolution. Such sampling requires experimental access to the continuously
ongoing dynamics of the network units. 

For a range of systems, however, direct access to the units' internal
states is not granted, but only times of events are available. Prominent
examples include the times of messages initiated or forwarded in online
social networks and distributed patterns of action potentials (spikes)
emitted by the neurons of brain circuits, both reflecting the respective
network structure in a non-trivial way \cite{Memmesheimer2006,Memmesheimer2006a,Jahnke:2008,Aoki2016,ICWSM1613026,Kirst2016}.
Reconstruction of physical network connectivity from such timing information
has been attempted for specific settings for neural circuits or online
social contacts. Often it is limited to small networks ($10^{2}$
units), by large computational efforts including high-performance
parallel computers up to about $10^{3}$ units, or to knowing specific
system models in advance \cite{Makarov:2005p9589,10.3389/fncom.2011.00003,Gerhard2013,Zaytsev2015,Aoki2016,ICWSM1613026,Cestnik2017}.
For instance, recent efforts on reconstructing spiking neural circuit
connectivity \cite{Zaytsev2015} show that combining stochastic mechanisms
for spike generation and linear kernels for spike integration enables
reconstruction of larger networks ($10^{3}$ neurons), if the spike
integration model closely matches the original simulated systems.
Alternatively, in systems of pulse-coupled units \cite{Cestnik2017}
reconstruction is feasible if the units are all intrinsic oscillators.
Besides such specific solutions, a general model-independent theory
based on timing information generated by the collective network dynamics
is as yet unknown.

In this Letter, we propose a general theory for reconstructing physical
network connectivity based only on the event timing patterns generated
by the collective spiking dynamics. The theory reveals existence and
absence of interactions, their activating or deactivating nature,
and enables reliable network reconstruction from regular as well as
irregular timing patterns, even if some (hidden) units cannot be observed.
The proposed reconstruction theory is model-independent (thereby,
purely data-driven) and trivially parallelizable because linearized
mappings for different units are computationally independent of each
other.

\begin{figure}
\begin{centering}
\includegraphics[scale=0.215]{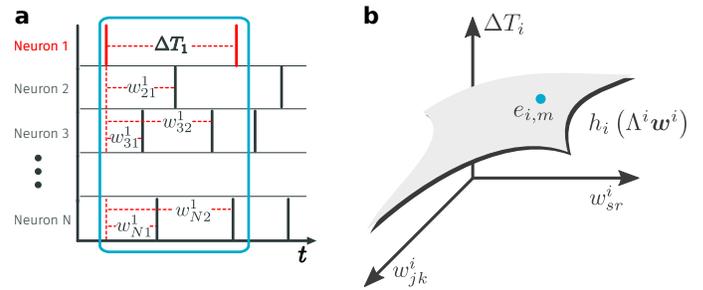}
\par\end{centering}
\caption{\textbf{Representing timing patterns in }\textbf{\emph{event space}}\textbf{.
a}, Schematics of an event for a sample unit ($i=1$) formed by the
inter-event time of unit $1$ and all cross-event intervals that may
influence the subsequent event of unit $1$. An additional counting
index $m$ is dropped for simplicity \textbf{b}, In the event space
for neuron $i$, each event $e_{i,m}$ is represented by a point on
the manifold defined through (\ref{eq:nonlinear}). \label{fig:Events-and-event}}
\end{figure}

\emph{Mapping timing patterns to physical connections?}\textbf{ }To
present the proposed theory consistently, we focus on a setting (and
notation) of networks of spiking neurons. Alternate applications work
in qualitatively the same way and are discussed towards the end of
this article. Here the units are individual neurons, a physical connection
is a synapse from one neuron to another and the events observed for
each unit are the electrical action potentials or spikes emitted by
the neuron. Specifically, inter-event intervals are inter-spike intervals
(ISIs) and cross-event intervals cross-spike intervals (CSIs). The
theory stays unchanged (up to notation) for all kinds of event times
observed from and originally generated by any specific collective
network dynamics that are typically unknown but coordinated via the
network interactions. 

Thus, consider a network of $N$ units $i\in\left\{ 1,\ldots,N\right\} $
generating spatio-temporal spike patterns (Fig. \ref{fig:Events-and-event})
defined by the sets of times $t_{i,m}$, where $m\in\mathbb{N}$ counts
the spike times. An inter-spike interval (ISI) 

\begin{equation}
\Delta T_{i,m}:=t_{i,m}-t_{i,m-1}>0,
\end{equation}
measures the duration of time between two consecutive events, the
$(m-1)$-st and the $m$-th spike times $t_{i,m-1}$ and $t_{i,m}$
of neuron $i$. Similarly, the cross-spike intervals (CSIs) 
\begin{equation}
w_{j,k,m}^{i}:=t_{j,p}-t_{i,m-1}>0,
\end{equation}
measure the duration between the $p$-th spike generated by neuron
$j$ and the previous ($(m-1)$st) spike generated by $i$ with $t_{j,p}<t_{i,m}$
(all superscripts throughout this article denote indices, not powers).
We index the CSIs by integers $k\equiv k(t_{i,m-1},t_{j,p})$ starting
with $k=1$ for the first spike at $t_{j,p}=\min_{p'}\left\{ t_{j,p'}\,|\,t_{j,p'}>t_{i,m-1}\right\} $
of unit $j$ after $t_{i,m-1}$ and sequentially increasing $k$ by
one by counting through the sequence of $t_{j,p}$ forwards in time.
Figure \ref{fig:Events-and-event} illustrates this definition for
one given ISI, where the indices $m-1$ and $m$ are suppressed. Now
consider that some finite number $K^{i}$ of spike times (for each
of the other neurons $j$) preceding $t_{i,m}$ influences the time
$t_{i,m}$ in a relevant way.

As a core conceptual step, we propose that the ISIs of each neuron
$i$ are approximately given by some unknown, locally smooth function
$h_{i}:\mathbb{R}^{N\times K^{i}}\rightarrow\mathbb{R}$ of the $K^{i}$
relevant cross-spike intervals (see Figure 1b) such that 
\begin{equation}
\Delta T_{i,m}=h_{i}\left(\Lambda^{i}W_{m}^{i}\right)\label{eq:nonlinear}
\end{equation}
for all $m$. Here the explicit dependency matrix $\Lambda^{i}\in\left\{ 0,1\right\} ^{N\times N}$
, cf. \cite{Casadiego2017}, is a diagonal matrix (to be determined)
indicating whether there is a physically active (synaptic) connection
from neuron $j$ to $i$ ($\Lambda_{jj}^{i}=1$) or not ($\Lambda_{jj}^{i}=0$).
The matrix 
\begin{equation}
W_{m}^{i}:=\left[\begin{array}{cccc}
w_{1,1,m}^{i} & w_{1,2,m}^{i} & \ldots & w_{1,K^{i},m}^{i}\\
w_{2,1,m}^{i} & w_{2,2,m}^{i} & \ldots & w_{2,K^{i},m}^{i}\\
\vdots & \vdots & \ddots & \vdots\\
w_{N,1,m}^{i} & w_{N,2,m}^{i} & \ldots & w_{N,K^{i},m}^{i}
\end{array}\right]\in\mathbb{R}^{N\times K^{i}},\label{eq:W}
\end{equation}
collects the $K^{i}$ CSIs generated by each neuron $j$ until just
before the $m$-th spike generated by neuron $i$. We refer to the
$k$-th column 
\begin{equation}
\boldsymbol{w}_{k,m}^{i}:=\left[w_{1,k,m}^{i},w_{2,k,m}^{i},\ldots,w_{N,k,m}^{i}\right]^{\mathsf{T}}\in\mathbb{R}^{N},\label{eq:firing_profile}
\end{equation}
of $W_{m}^{i}$ as the $k$-th \emph{presynaptic profile} of neuron
$i$ before its $m$-th spike time. It indicates when presynaptic
neurons fired for the first time, second time, and so on until $K^{i}$-th
time, before $t_{i,m}$. For illustration purposes, we here specifically
constrain $K^{i}$ to take into account only those spike times within
the currently considered ISI such that $t_{j,p}\in[t_{i,m-1},t_{i,m}]$.
If a presynaptic neuron does not spike within this interval, its CSIs
are set to zero, $w_{j,k,m}^{i}:=0$, in (\ref{eq:W}). From a biological
perspective, the function $h_{i}$ is determined by the intrinsic
properties of neuron $i$ including its spike generation mechanism
as well as its pre- and postsynaptic processes. The function $h_{i}$
is in general unknown. 

In particular, equation (\ref{eq:nonlinear}) assigns a specific ISI
to a specific collection of timings of presynaptic inputs. Therefore,
we may represent such ISI-CSIs tuple in a higher dimensional \emph{event
space }$\mathcal{E}_{i}\subset\mathbb{R}^{(NK^{i}+1)}$, where each
realization of neuron $i$'s dynamics is given by events defined as
\begin{equation}
e_{i,m}:=\left[\mbox{vec}\left(W_{m}^{i}\right),\Delta T_{i,m}\right]^{\mathsf{T}}\in\mathbb{R}^{(NK^{i}+1)},\label{eq:event_general}
\end{equation}
where the operator $\mbox{vec}:\mathbb{R}^{S\times R}\rightarrow\mathbb{R}^{SR}$
stands for vectorization of a matrix and it transforms a matrix into
a column vector \cite{Magnus1988}.

\begin{figure}
\begin{centering}
\includegraphics[scale=0.215]{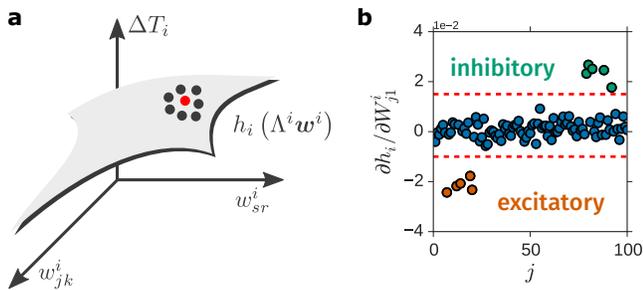}
\par\end{centering}
\caption{\textbf{Slopes in event space yield excitatory and inhibitory synaptic
interactions. a}, Schematics of a local sampling in the event space.
Local samplings may be taken from regular spiking patterns or subsets
of irregular spike patterns such that events are close-by in event
space. Taking a reference event (red dot) and linearly approximating
all other events (gray dot) through (\ref{eq:linearization_final})
constrains existence and sign of interactions without knowing a system
model. \textbf{b}, Reconstruction of inhibitory (green), excitatory
(orange) and absent (blue) synaptic interactions of a neuron in a
random network of $N=100$ LIF neurons having $\delta$-synapses
with $N_{exc}=50$ excitatory and, $N_{inh}=50$ inhibitory neurons
and a connection probability $p=0.1$. See \cite{Supp_Neuron}
for further parameters. The red-dashed lines indicate
the optimal thresholds (as calculated via Otsu's method \cite{Otsu1979})
to distinguish excitatory and inhibitory from absent interactions.
\label{fig:local}}
\end{figure}

So, how can events $e_{i,m}$ in the event space $\mathcal{E}_{i}$
help us determine the synaptic inputs to neuron $i$? Consider a local
sample of $M+1$ events in $\mathcal{E}_{i}$ as follows (Figure \ref{fig:local}a).
Selecting a reference event 
\begin{equation}
e_{i,r}:=\arg\min_{e_{i,s}}\sum_{m}\left\Vert e_{i,m}-e_{i,s}\right\Vert _{2},\label{eq:referenceEvent}
\end{equation}
 closest to all other events in the sample, with $m,s\in\left\{ 1,2\ldots,M+1\right\} $,
yields an approximation

\begin{equation}
\Delta T_{i,m}\doteq\Delta T_{i,r}+\mbox{tr}\left(\left(\dfrac{\partial h_{i}}{\partial W^{i}}\right)^{\mathsf{T}}\Lambda^{i}\left[W_{m}^{i}-W_{r}^{i}\right]\right),\label{eq:linearization}
\end{equation}
of (\ref{eq:nonlinear}) linear in the differences $\left[W_{m}^{i}-W_{r}^{i}\right]$
around $e_{i,r\,}$. Here $\left(\partial h_{i}/\partial W^{i}\right)\equiv\left(\partial h_{i}/\partial W^{i}\right)\left(\Lambda^{i}\left(W_{r}^{i}\right)\right)\in\mathbb{R}^{N\times K^{i}}$
is a matrix derivative and $\mbox{tr}(\cdot)$ is the trace operation.
Rewriting (\ref{eq:linearization}) yields
\begin{equation}
\Delta T_{i,m}\doteq\Delta T_{i,r}+\sum_{k=1}^{K_{i}}\boldsymbol{\nabla}h_{i,k}\Lambda^{i}\left[\boldsymbol{w}_{k,m}^{i}-\boldsymbol{w}_{k,r}^{i}\right],\label{eq:linearization_final}
\end{equation}
where $\boldsymbol{\nabla}h_{i,k}:=\left[\dfrac{\partial h_{i}}{\partial W_{1k}^{i}},\dfrac{\partial h_{i}}{\partial W_{2k}^{i}},\ldots,\dfrac{\partial h_{i}}{\partial W_{Nk}^{i}}\right]\in\mathbb{R}^{N}$
is the gradient of the function at the $k$-th presynaptic profile.
Thus, given $m\in\left\{ 1,2,\ldots,M\right\} $ different events
(in addition to the reference event $r$), finding the synaptic connectivity
becomes solving the linear regression problem (\ref{eq:linearization_final})
for the unknown parameters $\boldsymbol{\nabla}h_{i,k}\Lambda^{i}$.
In particular, system (\ref{eq:linearization_final}) may be computationally
solved in parallel for different units $i$. We here solve such linear
system via least squares minimization, compare \cite{Shandilya2011,Timme2014}.
Block-sparse regression algorithms may also be employed, especially
if one proposes higher-order approximations than (\ref{eq:linearization_final})
\cite{Eldar2009,Casadiego2017}.

\begin{figure}[t]
\begin{centering}
\includegraphics[scale=0.27]{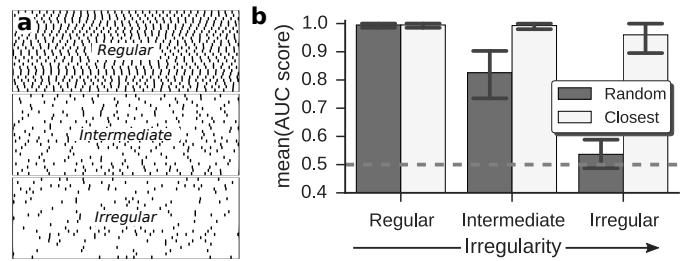}
\par\end{centering}
\caption{\textbf{Revealing synaptic connections from regular to irregular spiking
patterns.} \textbf{a}, Schematic representation of
spike trains with different degrees of irregularity. To establish
test statistics, the spike trains were sampled from blue random
networks of $N=100$ inhibitory LIF neurons interacting via $\delta$-synapses
with connection probability $p=0.1$. All simulations
were performed using identical time intervals of $500\,\text{s}$
each (see \cite{Supp_Neuron} for further settings). \textbf{b},
AUC scores for two different sampling conditions: random and closest.
In the random sampling paradigm, events are randomly drawn from the
uniform distribution across the spike trains, while in a closest sampling
paradigm, the same number of events closest to a reference event are
jointly considered for reconstruction. Reconstruction quality decreases
with spiking pattern irregularity for random sampling (dark gray)
yet stays consistently high for closest sampling. Gray dashed line
indicates random guessing. \label{fig:irregular}}
\end{figure}

\begin{figure}
\begin{centering}
\includegraphics[scale=0.2]{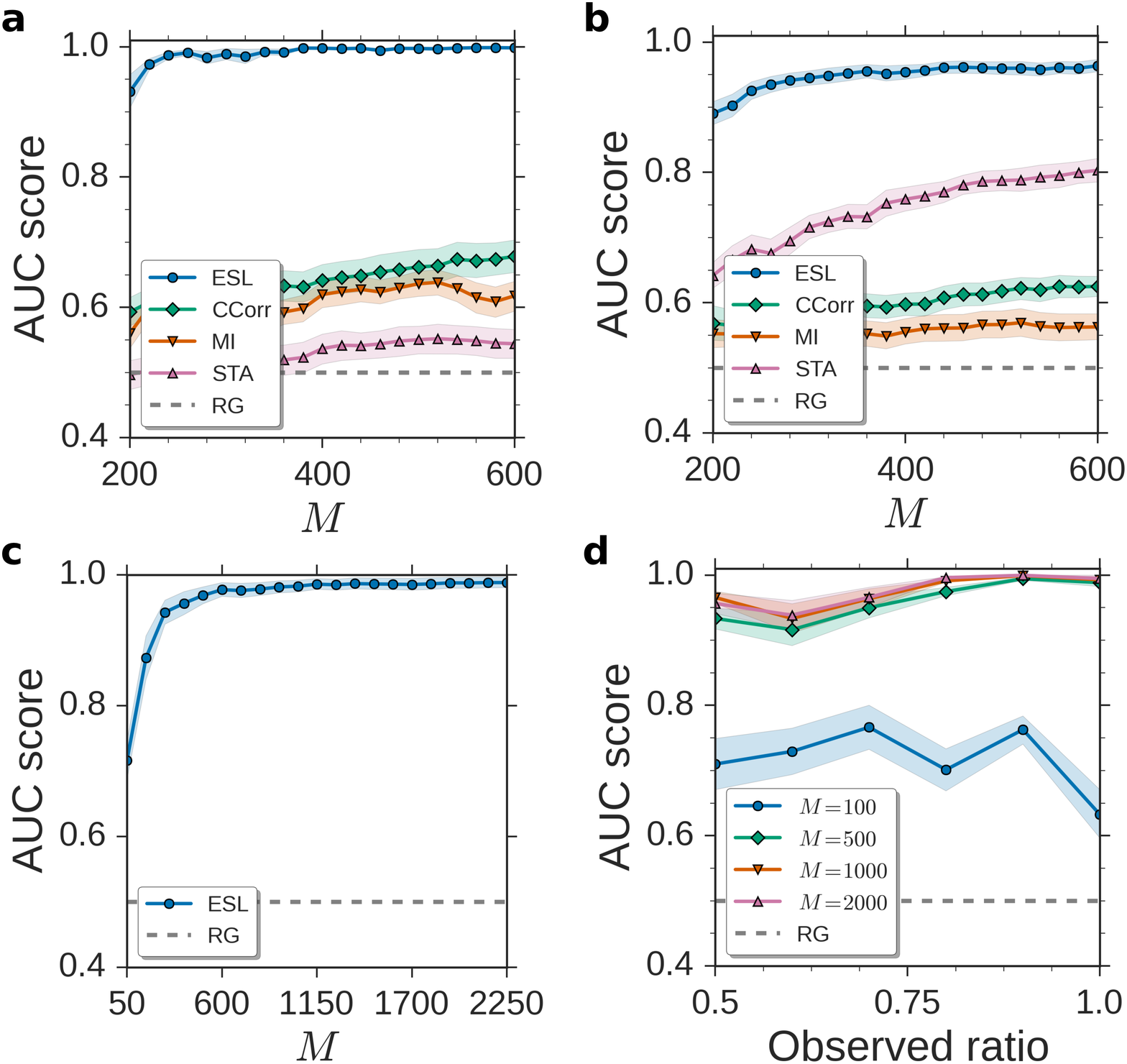}
\par\end{centering}
\caption{\textbf{Model-independence and robustness against hidden units. }Quality
of reconstruction vs. the number of events $M$ considered for blue random
networks of $N=100$ blue and $p=0.1$\textbf{ a}, LIF
neurons with $\alpha$-synapses (ESL, CCorr, MI, STA and RG stand
for event-space linearization (our approach, marked in blue), cross-correlations
(green), mutual information (orange), spike-triggered average (rose)
and random guessing (gray), respectively); \textbf{b}, Hodgkin-Huxley
neurons with conductance-based synapses; \textbf{c,} LIF neurons coupled
with $\delta$-synapses where only 80 neurons are observed.\textbf{
d}, Systematic reconstructions versus the fraction of observed neurons.
Network parameters are $N_{exc}=50$ and $N_{inh}=50$, see \cite{Supp_Neuron}
for further settings. \label{fig:models}}
\end{figure}

\emph{Revealing synapses in networks of spiking neurons.}\textbf{
}To validate the predictive power of our theory, we revealed the synaptic
connectivity (i) of model systems with current-based and conductance-based
synapses, (ii) of excitatory and inhibitory synapses, (iii) with both
instantaneous and temporally extended responses, (iv) of moderately
larger number of neurons (compared to the state-of-the-art), (v) for
regular and irregular spiking patterns and (vi) under conditions where
a subset of units is hidden or unavailable to observation (Figures
\ref{fig:local}, \ref{fig:irregular}, and \ref{fig:models}). We
quantify performance of reconstruction by the Area Under the Receiver-Operating-Characteristic
Curve (AUC) score \cite{Fawcett2006}, that equals $1$ for perfect
reconstruction and $1/2$ for predictions as good as random guessing. 

We start illustrating successful reconstruction for simple networks
of leaky integrate-and-fire (LIF) \cite{Brunel} neurons with mixed
inhibitory and excitatory current-based $\delta$-synapses (see Supplemental
Material \cite{Supp_Neuron}). As Figure \ref{fig:local}b shows by
example, the method of event-space mappings does not only reveal the
existence and absence of synapses between pairs of neurons in the
network; in particular, the signs and magnitudes of the derivatives
$\left(\partial h_{i}/\partial W_{j1}^{i}\right)$ already indicate
whether a synapse acts in an inhibitory or excitatory way: indeed,
inhibitory inputs to neuron $i$ retard its subsequent spike and thereby
extend the duration of an ISI, whereas excitatory inputs shorten the
duration. Thus, positive derivatives indicate (effectively) inhibitory
and negative derivatives excitatory interactions, compare Figure \ref{fig:local}b.

\emph{Reconstruction from irregular event patterns?} What if the spiking
patterns are not regular as often the case for real data (from neurophysiological
recordings as well as from observations in any other discipline) and
thus not all events are located sufficiently close to one reference
event (\ref{eq:referenceEvent})? If the induced events $e_{i,m}$
are distributed in event space $\mathcal{E}_{i}$ less locally, with
the $e_{i,m}$ located on different, possibly non-adjacent patches
in $\mathcal{E}_{i}$, we systematically collect only those events
that are located close to selected reference events $e_{i,r\,}$\footnote{One may alternatively collect events close to several references $e_{i,r\,}$,
$e_{i,r'\,}$, etc. and concatenate local (and generally different)
approximations of the form (\ref{eq:linearization_final}) yielding
constraints of the same (linear) type and solve the regression problem
via block-sparse regression \cite{Casadiego2017}.}. To check performance, we systematically varied how irregular spike
timing patterns are by varying the overall coupling strength by a
factor of 30, effectively interpolating between regimes close to regular
dynamics (locking) \cite{Vreeswijk1996} for small coupling and close
to irregular balanced states \cite{VanVreeswijk1996,Timme:2002,Jahnke:2008,Hansel2013}
for large coupling. If sufficiently many events (ISI-CSI combinations)
that are close in event space occur in the network's spike timing
patterns, sampling the events in some local region (or, alternatively,
local regions) in event space may be compensated by longer recording
times or collecting several patches of events. With
increasing inhibition, the total number of spikes in each recording
decreases whereas the spike sequence irregularity increases. Thereby,
to ensure a sufficient number of events locally in event space across
inhibition levels, we simulated all networks for 500 s each. We indeed
find that such closest sampling paradigm enables consistent high-quality
reconstruction for regular, intermediate and irregular spike timing
patterns alike (Fig. \ref{fig:irregular}).

We reiterate that approximation (\ref{eq:linearization_final}) requires
no \emph{a priori} information about circuit or neuron models and
parameters, instead, only spike timing data are necessary to reveal
synapses. The same event-space/linearization (ESL) method proposed
performs robustly across different circuits of neurons (low- and higher-dimensional
neuron models and different synaptic models), compares favorably to
alternative approaches of model-free reconstruction and even yields
reasonable estimates if some units are hidden (see
\cite{Supp_Neuron} for more details), see Figure \ref{fig:models}.
Figure panels \ref{fig:models}a and b illustrate the performance
for networks of simple LIF neurons with current-based synapses as
well as for more biophysically detailed Hodgkin-Huxley neurons coupled
via conductance-based synapses (see\textbf{ }\cite{Supp_Neuron})
explicating that the theory is insensitive to changes of types of
coupling and neurons. The ESL method outperforms predictions by statistical
dependencies such as cross-correlations, mutual information and spike-triggered
averages. In addition, if some neurons are inaccessible (hidden units),
synapses among accessible neurons may still be recovered (see
\cite{Supp_Neuron}), Figure \ref{fig:models}c-d. Furthermore,
a systematic study shows that ESL accurately determines synaptic links
in the presence of external inputs emulating neurons firing with Poisson
statistics, at the expense of requiring to record a larger number
of events (see Supplementary Material \cite{Supp_Neuron} for a detailed
study). As explained above, recording a larger number of events promotes
denser samplings in the event space, which in turn aids in filtering
out dynamical effects related to such unobserved inputs when solving
the regression problem.

\begin{figure}
\begin{centering}
\includegraphics[scale=0.18]{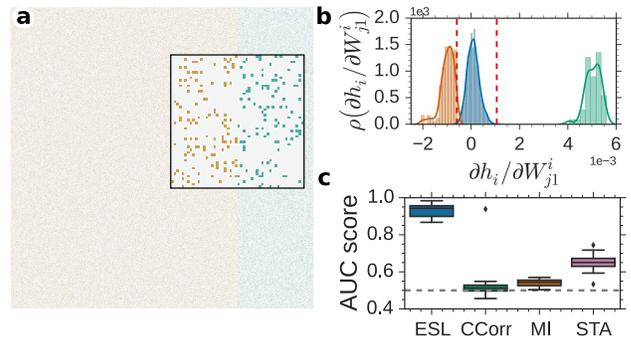}
\par\end{centering}
\caption{\textbf{Reconstructing large networks is computationally feasible.}
Reconstruction of random networks of $N=2000$ LIF
neurons with $\delta$-synapses and $p=0.1$. See
\cite{Supp_Neuron} for more details. \textbf{a}, Connectivity matrix
excitatory (red) and inhibitory (blue) synapses. Inset shows a close
up. \textbf{b}, Distribution of reconstructed excitatory (orange),
absent (blue) and inhibitory (green) synapses of a single postsynaptic
neuron. \textbf{c}, Quality of reconstruction for individual neurons
for $M=8000$. ESL, CCorr, MI and STA stand for the approaches introduced
here, using cross-correlations, mutual information and spike-triggered
averages, respectively. The gray-dashed line stands for random guessing.
\label{fig:large}}
\end{figure}

Finally, reconstructing larger networks seems computationally feasible
and reconstruction quality compares favorably to other general, model-free
approaches. For instance, Figure \ref{fig:large} illustrates successful
reconstruction of the presynaptic pattern of a random unit $i$ in
a network of $N=2000$ neurons exhibiting regular dynamics that was
computed within about 500 seconds ($\sim10$ minutes per neuron) on
a single machine \footnote{Reconstructions were performed on a single core of a desktop computer
with Intel\textregistered{} Core\texttrademark{} i7-4770 CPU@3.40GHz
octocore processor and 16 GB of RAM memory . Most of execution time
is exhausted in the computation of distances among the events. Reconstruction
time is further determined by the speed of the numerical package chosen
to perform least squares optimization. }. Generally, the computational (time) complexity of our reconstruction
theory resutls from the computation of event-space distances, which
scales as $\mathcal{O}\left((MNK)^{2}\right)$ (see \cite{Supp_Neuron}).
The approach separates not only existing from absent but also excitatory
from inhibitory synapses. As before, a systematic comparison for equal
number of events shows that our approach again outperforms predictions
by commonly employed statistical dependency quantifiers, Figure \ref{fig:large}c.

\emph{Summary and Conclusions.}\textbf{ }We presented a general, model-independent
theory for reconstructing the topology of a network's direct interactions
from observed patterns of event timings only. A key advance is the
proposal that the interevent intervals are given by some unknown,
sufficiently smooth function, thereby not requiring access to a event
generating system model or even the full state vector of the dynamical
system generating the events. We illustrated core performance aspects
including robustness (against changes in unit dynamics and coupling
schemes), computational performance (rapid analysis on single machines
and parallelizability) and suitability even if some units are hidden
by example of spiking neural networks. By representing inter-event
intervals as functions of cross-event intervals, we mapped the problem
to event spaces yielding linear equations that enable robust least
squares solutions for the topology. Thereby, the approach may be transferred
to other systems generating event time series. 

We remark that $\boldsymbol{\nabla}h_{i,k}\Lambda^{i}$ in (\ref{eq:linearization_final})
maps\emph{ }exactly those physical connections that are used for transmitting
signals that actually directly influence the timings $t_{i,m}$ and
thus the inter-event times. This has distinct consequences in practice.
For instance, a presynaptic neuron $j$ may generate its only potentially
relevant spike during the refractory period of neuron $i$ (or may
simply not generate any spike) during the observation period of the
experiment recording the timing pattern, thus no synapse from unit
$j$ will be indicated even if an anatomical one exists. At the same
time, if for any reason (e.g. measurement error or data corruption)
an event of unit $i$ is not recorded at all or the ISI recorded with
a large error, it becomes easy to spot this event as an outlier as
it would lie far above or below the other events with nearby CSIs.
As only the presence (and sign) of the above derivatives play a role
for reconstruction, a straightforward generalization is to collect
events close to several references $e_{i,r\,}$, $e_{i,r'\,}$, etc.
and concatenate local (and generally different) approximations of
the form (\ref{eq:linearization_final}).

A recent work \cite{English2017} studying excitatory-to-inhibitory
CA1 synapses \emph{in vivo }focused
on predicting exclusively the strongest interactions from excitatory
to inhibitory neurons using a statistical GLM-based method combined
with cross-correlograms, which results in high-computational demands
when applied on large networks. The clearest advantages
of our theory, beyond its simplicity, are its rapid computational
performance and its generality and model-independence, compare to
a recent alternative approach \cite{Cestnik2017}. Moreover, our ansatz
by construction generalizes to systems beyond spiking neural networks
(used here as a systematic case study for illustration) and may be
applied, for instance, for revealing friendship networks in online
social networks \cite{Aoki2016,ICWSM1613026} or to determine who
communicates with whom in hidden web or wireless services \cite{Bettstetter2002,Klinglmayr2012}
\textendash{} from the timing of local unit activities alone. Taken
together, the results illustrated above suggest the power of systematically
generalizing a state space perspective for representing the full trajectories
to an event space perspective for representing collectively coordinated
event timing patterns.

\emph{Acknowledgements.} We thank Viola Priesemann, Juan Florez-Weidinger,
Mantas Ga\-brie\-lai\-tis and Stayko Popov for valuable discussions.
We gratefully acknowledge support from the Federal Ministry of Education
and Research (BMBF Grant No. 03SF0472F), the Max Planck Society and
the German Research Foundation (DFG) through the Cluster of Excellence
\emph{Center for Advancing Electronics Dresden} (cfaed). J. Casadiego
and D. Maoutsa contributed equally to this work.

\bibliographystyle{apsrev4-1}

\begin{thebibliography}{37}%
\makeatletter
\providecommand \@ifxundefined [1]{%
 \@ifx{#1\undefined}
}%
\providecommand \@ifnum [1]{%
 \ifnum #1\expandafter \@firstoftwo
 \else \expandafter \@secondoftwo
 \fi
}%
\providecommand \@ifx [1]{%
 \ifx #1\expandafter \@firstoftwo
 \else \expandafter \@secondoftwo
 \fi
}%
\providecommand \natexlab [1]{#1}%
\providecommand \enquote  [1]{``#1''}%
\providecommand \bibnamefont  [1]{#1}%
\providecommand \bibfnamefont [1]{#1}%
\providecommand \citenamefont [1]{#1}%
\providecommand \href@noop [0]{\@secondoftwo}%
\providecommand \href [0]{\begingroup \@sanitize@url \@href}%
\providecommand \@href[1]{\@@startlink{#1}\@@href}%
\providecommand \@@href[1]{\endgroup#1\@@endlink}%
\providecommand \@sanitize@url [0]{\catcode `\\12\catcode `\$12\catcode
  `\&12\catcode `\#12\catcode `\^12\catcode `\_12\catcode `\%12\relax}%
\providecommand \@@startlink[1]{}%
\providecommand \@@endlink[0]{}%
\providecommand \url  [0]{\begingroup\@sanitize@url \@url }%
\providecommand \@url [1]{\endgroup\@href {#1}{\urlprefix }}%
\providecommand \urlprefix  [0]{URL }%
\providecommand \Eprint [0]{\href }%
\providecommand \doibase [0]{http://dx.doi.org/}%
\providecommand \selectlanguage [0]{\@gobble}%
\providecommand \bibinfo  [0]{\@secondoftwo}%
\providecommand \bibfield  [0]{\@secondoftwo}%
\providecommand \translation [1]{[#1]}%
\providecommand \BibitemOpen [0]{}%
\providecommand \bibitemStop [0]{}%
\providecommand \bibitemNoStop [0]{.\EOS\space}%
\providecommand \EOS [0]{\spacefactor3000\relax}%
\providecommand \BibitemShut  [1]{\csname bibitem#1\endcsname}%
\let\auto@bib@innerbib\@empty
\bibitem [{\citenamefont {Yeung}\ \emph {et~al.}(2002)\citenamefont {Yeung},
  \citenamefont {Tegner},\ and\ \citenamefont {Collins}}]{Yeung:2002}%
  \BibitemOpen
  \bibfield  {author} {\bibinfo {author} {\bibfnamefont {M.~K.}\ \bibnamefont
  {Yeung}}, \bibinfo {author} {\bibfnamefont {J.}~\bibnamefont {Tegner}}, \
  and\ \bibinfo {author} {\bibfnamefont {J.~J.}\ \bibnamefont {Collins}},\
  }\href {\doibase 10.1073/pnas.092576199} {\bibfield  {journal} {\bibinfo
  {journal} {Proc. Natl. Acad. Sci. U. S. A.}\ }\textbf {\bibinfo {volume}
  {99}},\ \bibinfo {pages} {6163} (\bibinfo {year} {2002})}\BibitemShut
  {NoStop}%
\bibitem [{\citenamefont {Gardner}\ \emph {et~al.}(2003)\citenamefont
  {Gardner}, \citenamefont {di~Bernardo}, \citenamefont {Lorenz},\ and\
  \citenamefont {Collins}}]{Gardner2003}%
  \BibitemOpen
  \bibfield  {author} {\bibinfo {author} {\bibfnamefont {T.~S.}\ \bibnamefont
  {Gardner}}, \bibinfo {author} {\bibfnamefont {D.}~\bibnamefont
  {di~Bernardo}}, \bibinfo {author} {\bibfnamefont {D.}~\bibnamefont {Lorenz}},
  \ and\ \bibinfo {author} {\bibfnamefont {J.~J.}\ \bibnamefont {Collins}},\
  }\href {\doibase 10.1126/science.1081900} {\bibfield  {journal} {\bibinfo
  {journal} {Science}\ }\textbf {\bibinfo {volume} {301}},\ \bibinfo {pages}
  {102} (\bibinfo {year} {2003})}\BibitemShut {NoStop}%
\bibitem [{\citenamefont {Yu}\ \emph {et~al.}(2006)\citenamefont {Yu},
  \citenamefont {Righero},\ and\ \citenamefont {Kocarev}}]{Yu2006}%
  \BibitemOpen
  \bibfield  {author} {\bibinfo {author} {\bibfnamefont {D.}~\bibnamefont
  {Yu}}, \bibinfo {author} {\bibfnamefont {M.}~\bibnamefont {Righero}}, \ and\
  \bibinfo {author} {\bibfnamefont {L.}~\bibnamefont {Kocarev}},\ }\href
  {\doibase 10.1103/PhysRevLett.97.188701} {\bibfield  {journal} {\bibinfo
  {journal} {Phys. Rev. Lett.}\ }\textbf {\bibinfo {volume} {97}},\ \bibinfo
  {pages} {188701} (\bibinfo {year} {2006})}\BibitemShut {NoStop}%
\bibitem [{\citenamefont {Timme}(2007)}]{Timme2007}%
  \BibitemOpen
  \bibfield  {author} {\bibinfo {author} {\bibfnamefont {M.}~\bibnamefont
  {Timme}},\ }\href@noop {} {\bibfield  {journal} {\bibinfo  {journal} {Phys.
  Rev. Lett.}\ }\textbf {\bibinfo {volume} {98}},\ \bibinfo {pages} {224101}
  (\bibinfo {year} {2007})}\BibitemShut {NoStop}%
\bibitem [{\citenamefont {Yu}\ and\ \citenamefont {Parlitz}(2010)}]{Yu:2010a}%
  \BibitemOpen
  \bibfield  {author} {\bibinfo {author} {\bibfnamefont {D.}~\bibnamefont
  {Yu}}\ and\ \bibinfo {author} {\bibfnamefont {U.}~\bibnamefont {Parlitz}},\
  }\href {\doibase 10.1103/PhysRevE.82.026108} {\bibfield  {journal} {\bibinfo
  {journal} {Phys. Rev. E}\ }\textbf {\bibinfo {volume} {82}},\ \bibinfo
  {pages} {026108} (\bibinfo {year} {2010})}\BibitemShut {NoStop}%
\bibitem [{\citenamefont {Wang}\ \emph {et~al.}(2016)\citenamefont {Wang},
  \citenamefont {Lai},\ and\ \citenamefont {Grebogi}}]{Wang2016}%
  \BibitemOpen
  \bibfield  {author} {\bibinfo {author} {\bibfnamefont {W.-X.}\ \bibnamefont
  {Wang}}, \bibinfo {author} {\bibfnamefont {Y.-C.}\ \bibnamefont {Lai}}, \
  and\ \bibinfo {author} {\bibfnamefont {C.}~\bibnamefont {Grebogi}},\ }\href
  {\doibase 10.1016/j.physrep.2016.06.004} {\bibfield  {journal} {\bibinfo
  {journal} {Phys. Rep.}\ }\textbf {\bibinfo {volume} {644}},\ \bibinfo {pages}
  {1} (\bibinfo {year} {2016})}\BibitemShut {NoStop}%
\bibitem [{\citenamefont {Barzel}\ and\ \citenamefont
  {Barab{\'{a}}si}(2013)}]{Barzel:2013}%
  \BibitemOpen
  \bibfield  {author} {\bibinfo {author} {\bibfnamefont {B.}~\bibnamefont
  {Barzel}}\ and\ \bibinfo {author} {\bibfnamefont {A.~L.}\ \bibnamefont
  {Barab{\'{a}}si}},\ }\href@noop {} {\bibfield  {journal} {\bibinfo  {journal}
  {Nat. Biotechnol.}\ }\textbf {\bibinfo {volume} {31}},\ \bibinfo {pages}
  {720} (\bibinfo {year} {2013})}\BibitemShut {NoStop}%
\bibitem [{\citenamefont {Timme}\ and\ \citenamefont
  {Casadiego}(2014)}]{Timme2014}%
  \BibitemOpen
  \bibfield  {author} {\bibinfo {author} {\bibfnamefont {M.}~\bibnamefont
  {Timme}}\ and\ \bibinfo {author} {\bibfnamefont {J.}~\bibnamefont
  {Casadiego}},\ }\href {\doibase 10.1088/1751-8113/47/34/343001} {\bibfield
  {journal} {\bibinfo  {journal} {J. Phys. A Math. Theor.}\ }\textbf {\bibinfo
  {volume} {47}},\ \bibinfo {pages} {343001} (\bibinfo {year} {2014})},\
  \Eprint {http://arxiv.org/abs/1408.2963} {1408.2963} \BibitemShut {NoStop}%
\bibitem [{\citenamefont {Nitzan}\ \emph {et~al.}(2017)\citenamefont {Nitzan},
  \citenamefont {Casadiego},\ and\ \citenamefont {Timme}}]{Nitzan2017}%
  \BibitemOpen
  \bibfield  {author} {\bibinfo {author} {\bibfnamefont {M.}~\bibnamefont
  {Nitzan}}, \bibinfo {author} {\bibfnamefont {J.}~\bibnamefont {Casadiego}}, \
  and\ \bibinfo {author} {\bibfnamefont {M.}~\bibnamefont {Timme}},\ }\href
  {http://advances.sciencemag.org/content/3/2/e1600396} {\bibfield  {journal}
  {\bibinfo  {journal} {Sci. Adv.}\ }\textbf {\bibinfo {volume} {3}},\ \bibinfo
  {pages} {e1600396} (\bibinfo {year} {2017})}\BibitemShut {NoStop}%
\bibitem [{\citenamefont {Memmesheimer}\ and\ \citenamefont
  {Timme}(2006{\natexlab{a}})}]{Memmesheimer2006}%
  \BibitemOpen
  \bibfield  {author} {\bibinfo {author} {\bibfnamefont {R.~M.}\ \bibnamefont
  {Memmesheimer}}\ and\ \bibinfo {author} {\bibfnamefont {M.}~\bibnamefont
  {Timme}},\ }\href@noop {} {\bibfield  {journal} {\bibinfo  {journal} {Physica
  D}\ }\textbf {\bibinfo {volume} {224}},\ \bibinfo {pages} {182} (\bibinfo
  {year} {2006}{\natexlab{a}})}\BibitemShut {NoStop}%
\bibitem [{\citenamefont {Memmesheimer}\ and\ \citenamefont
  {Timme}(2006{\natexlab{b}})}]{Memmesheimer2006a}%
  \BibitemOpen
  \bibfield  {author} {\bibinfo {author} {\bibfnamefont {R.~M.}\ \bibnamefont
  {Memmesheimer}}\ and\ \bibinfo {author} {\bibfnamefont {M.}~\bibnamefont
  {Timme}},\ }\href@noop {} {\bibfield  {journal} {\bibinfo  {journal} {Phys.
  Rev. Lett.}\ }\textbf {\bibinfo {volume} {97}},\ \bibinfo {pages} {188101}
  (\bibinfo {year} {2006}{\natexlab{b}})}\BibitemShut {NoStop}%
\bibitem [{\citenamefont {Jahnke}\ \emph {et~al.}(2008)\citenamefont {Jahnke},
  \citenamefont {Memmesheimer},\ and\ \citenamefont {Timme}}]{Jahnke:2008}%
  \BibitemOpen
  \bibfield  {author} {\bibinfo {author} {\bibfnamefont {S.}~\bibnamefont
  {Jahnke}}, \bibinfo {author} {\bibfnamefont {R.~M.}\ \bibnamefont
  {Memmesheimer}}, \ and\ \bibinfo {author} {\bibfnamefont {M.}~\bibnamefont
  {Timme}},\ }\href@noop {} {\bibfield  {journal} {\bibinfo  {journal} {Phys.
  Rev. Lett.}\ }\textbf {\bibinfo {volume} {100}},\ \bibinfo {pages} {048102}
  (\bibinfo {year} {2008})}\BibitemShut {NoStop}%
\bibitem [{\citenamefont {Aoki}\ \emph {et~al.}(2016)\citenamefont {Aoki},
  \citenamefont {Takaguchi}, \citenamefont {Kobayashi},\ and\ \citenamefont
  {Lambiotte}}]{Aoki2016}%
  \BibitemOpen
  \bibfield  {author} {\bibinfo {author} {\bibfnamefont {T.}~\bibnamefont
  {Aoki}}, \bibinfo {author} {\bibfnamefont {T.}~\bibnamefont {Takaguchi}},
  \bibinfo {author} {\bibfnamefont {R.}~\bibnamefont {Kobayashi}}, \ and\
  \bibinfo {author} {\bibfnamefont {R.}~\bibnamefont {Lambiotte}},\ }\href
  {\doibase 10.1103/PhysRevE.94.042313} {\bibfield  {journal} {\bibinfo
  {journal} {Phys. Rev. E}\ }\textbf {\bibinfo {volume} {94}},\ \bibinfo
  {pages} {042313} (\bibinfo {year} {2016})}\BibitemShut {NoStop}%
\bibitem [{\citenamefont {Kobayashi}\ and\ \citenamefont
  {Lambiotte}(2016)}]{ICWSM1613026}%
  \BibitemOpen
  \bibfield  {author} {\bibinfo {author} {\bibfnamefont {R.}~\bibnamefont
  {Kobayashi}}\ and\ \bibinfo {author} {\bibfnamefont {R.}~\bibnamefont
  {Lambiotte}},\ }in\ \href
  {https://www.aaai.org/ocs/index.php/ICWSM/ICWSM16/paper/view/13026} {\emph
  {\bibinfo {booktitle} {Tenth International AAAI Conference on Web and Social
  Media}}}\ (\bibinfo {year} {2016})\BibitemShut {NoStop}%
\bibitem [{\citenamefont {Kirst}\ \emph {et~al.}(2016)\citenamefont {Kirst},
  \citenamefont {Timme},\ and\ \citenamefont {Battaglia}}]{Kirst2016}%
  \BibitemOpen
  \bibfield  {author} {\bibinfo {author} {\bibfnamefont {C.}~\bibnamefont
  {Kirst}}, \bibinfo {author} {\bibfnamefont {M.}~\bibnamefont {Timme}}, \ and\
  \bibinfo {author} {\bibfnamefont {D.}~\bibnamefont {Battaglia}},\ }\href
  {\doibase 10.1038/ncomms11061} {\bibfield  {journal} {\bibinfo  {journal}
  {Nat. Commun.}\ }\textbf {\bibinfo {volume} {7}},\ \bibinfo {pages} {11061}
  (\bibinfo {year} {2016})}\BibitemShut {NoStop}%
\bibitem [{\citenamefont {Makarov}\ \emph {et~al.}(2005)\citenamefont
  {Makarov}, \citenamefont {Panetsos},\ and\ \citenamefont {{De
  Feo}}}]{Makarov:2005p9589}%
  \BibitemOpen
  \bibfield  {author} {\bibinfo {author} {\bibfnamefont {V.~A.}\ \bibnamefont
  {Makarov}}, \bibinfo {author} {\bibfnamefont {F.}~\bibnamefont {Panetsos}}, \
  and\ \bibinfo {author} {\bibfnamefont {O.}~\bibnamefont {{De Feo}}},\ }\href
  {\doibase 10.1016/j.jneumeth.2004.11.013} {\bibfield  {journal} {\bibinfo
  {journal} {J. Neurosci. Methods}\ }\textbf {\bibinfo {volume} {144}},\
  \bibinfo {pages} {265} (\bibinfo {year} {2005})}\BibitemShut {NoStop}%
\bibitem [{\citenamefont {{Van Bussel}}\ \emph {et~al.}(2011)\citenamefont
  {{Van Bussel}}, \citenamefont {Kriener},\ and\ \citenamefont
  {Timme}}]{10.3389/fncom.2011.00003}%
  \BibitemOpen
  \bibfield  {author} {\bibinfo {author} {\bibfnamefont {F.}~\bibnamefont {{Van
  Bussel}}}, \bibinfo {author} {\bibfnamefont {B.}~\bibnamefont {Kriener}}, \
  and\ \bibinfo {author} {\bibfnamefont {M.}~\bibnamefont {Timme}},\ }\href
  {\doibase 10.3389/fncom.2011.00003} {\bibfield  {journal} {\bibinfo
  {journal} {Front. Comput. Neurosci.}\ }\textbf {\bibinfo {volume} {5}},\
  \bibinfo {pages} {3} (\bibinfo {year} {2011})}\BibitemShut {NoStop}%
\bibitem [{\citenamefont {Gerhard}\ \emph {et~al.}(2013)\citenamefont
  {Gerhard}, \citenamefont {Kispersky}, \citenamefont {Gutierrez},
  \citenamefont {Marder}, \citenamefont {Kramer},\ and\ \citenamefont
  {Eden}}]{Gerhard2013}%
  \BibitemOpen
  \bibfield  {author} {\bibinfo {author} {\bibfnamefont {F.}~\bibnamefont
  {Gerhard}}, \bibinfo {author} {\bibfnamefont {T.}~\bibnamefont {Kispersky}},
  \bibinfo {author} {\bibfnamefont {G.~J.}\ \bibnamefont {Gutierrez}}, \bibinfo
  {author} {\bibfnamefont {E.}~\bibnamefont {Marder}}, \bibinfo {author}
  {\bibfnamefont {M.}~\bibnamefont {Kramer}}, \ and\ \bibinfo {author}
  {\bibfnamefont {U.}~\bibnamefont {Eden}},\ }\href {\doibase
  10.1371/journal.pcbi.1003138} {\bibfield  {journal} {\bibinfo  {journal}
  {PLoS Comput. Biol.}\ }\textbf {\bibinfo {volume} {9}},\ \bibinfo {pages}
  {e1003138} (\bibinfo {year} {2013})}\BibitemShut {NoStop}%
\bibitem [{\citenamefont {Zaytsev}\ \emph {et~al.}(2015)\citenamefont
  {Zaytsev}, \citenamefont {Morrison},\ and\ \citenamefont
  {Deger}}]{Zaytsev2015}%
  \BibitemOpen
  \bibfield  {author} {\bibinfo {author} {\bibfnamefont {Y.~V.}\ \bibnamefont
  {Zaytsev}}, \bibinfo {author} {\bibfnamefont {A.}~\bibnamefont {Morrison}}, \
  and\ \bibinfo {author} {\bibfnamefont {M.}~\bibnamefont {Deger}},\
  }\href@noop {} {\bibfield  {journal} {\bibinfo  {journal} {J. Comput.
  Neurosci.}\ }\textbf {\bibinfo {volume} {39}},\ \bibinfo {pages} {77}
  (\bibinfo {year} {2015})}\BibitemShut {NoStop}%
\bibitem [{\citenamefont {Cestnik}\ and\ \citenamefont
  {Rosenblum}(2017)}]{Cestnik2017}%
  \BibitemOpen
  \bibfield  {author} {\bibinfo {author} {\bibfnamefont {R.}~\bibnamefont
  {Cestnik}}\ and\ \bibinfo {author} {\bibfnamefont {M.}~\bibnamefont
  {Rosenblum}},\ }\href {http://arxiv.org/abs/1704.06224} {\bibfield  {journal}
  {\bibinfo  {journal} {Phys. Rev. E}\ }\textbf {\bibinfo {volume} {96}},\
  \bibinfo {pages} {012209} (\bibinfo {year} {2017})}\BibitemShut {NoStop}%
\bibitem [{\citenamefont {Casadiego}\ \emph {et~al.}(2017)\citenamefont
  {Casadiego}, \citenamefont {Nitzan}, \citenamefont {Hallerberg},\ and\
  \citenamefont {Timme}}]{Casadiego2017}%
  \BibitemOpen
  \bibfield  {author} {\bibinfo {author} {\bibfnamefont {J.}~\bibnamefont
  {Casadiego}}, \bibinfo {author} {\bibfnamefont {M.}~\bibnamefont {Nitzan}},
  \bibinfo {author} {\bibfnamefont {S.}~\bibnamefont {Hallerberg}}, \ and\
  \bibinfo {author} {\bibfnamefont {M.}~\bibnamefont {Timme}},\ }\href@noop {}
  {\bibfield  {journal} {\bibinfo  {journal} {Nat. Commun.}\ }\textbf {\bibinfo
  {volume} {8}},\ \bibinfo {pages} {2192} (\bibinfo {year} {2017})}\BibitemShut
  {NoStop}%
\bibitem [{\citenamefont {Magnus}\ and\ \citenamefont
  {Neudecker}(1989)}]{Magnus1988}%
  \BibitemOpen
  \bibfield  {author} {\bibinfo {author} {\bibfnamefont {J.}~\bibnamefont
  {Magnus}}\ and\ \bibinfo {author} {\bibfnamefont {H.}~\bibnamefont
  {Neudecker}},\ }\href {\doibase 10.2307/1270024} {\emph {\bibinfo {title}
  {Technometrics}}},\ Vol.~\bibinfo {volume} {31}\ (\bibinfo {year} {1989})\
  p.\ \bibinfo {pages} {501}\BibitemShut {NoStop}%
\bibitem [{\citenamefont {Shandilya}\ and\ \citenamefont
  {Timme}(2011)}]{Shandilya2011}%
  \BibitemOpen
  \bibfield  {author} {\bibinfo {author} {\bibfnamefont {S.~G.}\ \bibnamefont
  {Shandilya}}\ and\ \bibinfo {author} {\bibfnamefont {M.}~\bibnamefont
  {Timme}},\ }\href {\doibase 10.1088/1367-2630/13/1/013004} {\bibfield
  {journal} {\bibinfo  {journal} {New J. Phys.}\ }\textbf {\bibinfo {volume}
  {13}},\ \bibinfo {pages} {013004} (\bibinfo {year} {2011})}\BibitemShut
  {NoStop}%
\bibitem [{\citenamefont {Eldar}\ and\ \citenamefont
  {Mishali}(2009)}]{Eldar2009}%
  \BibitemOpen
  \bibfield  {author} {\bibinfo {author} {\bibfnamefont {Y.~C.}\ \bibnamefont
  {Eldar}}\ and\ \bibinfo {author} {\bibfnamefont {M.}~\bibnamefont
  {Mishali}},\ }\href {\doibase 10.1109/TIT.2009.2030471} {\bibfield  {journal}
  {\bibinfo  {journal} {IEEE Trans. Inf. Theory}\ }\textbf {\bibinfo {volume}
  {55}},\ \bibinfo {pages} {5302} (\bibinfo {year} {2009})}\BibitemShut
  {NoStop}%
\bibitem [{\citenamefont {Fawcett}(2006)}]{Fawcett2006}%
  \BibitemOpen
  \bibfield  {author} {\bibinfo {author} {\bibfnamefont {T.}~\bibnamefont
  {Fawcett}},\ }\href {\doibase 10.1016/j.patrec.2005.10.010} {\bibfield
  {journal} {\bibinfo  {journal} {Pattern Recognit. Lett.}\ }\textbf {\bibinfo
  {volume} {27}},\ \bibinfo {pages} {861} (\bibinfo {year} {2006})}\BibitemShut
  {NoStop}%
\bibitem [{Sup()}]{Supp_Neuron}%
  \BibitemOpen
  \href@noop {} {\bibinfo  {journal} {See Supplemental Material for more
  details on neuron network models, simulation parameters and additional
  results}\ }\BibitemShut {NoStop}%
\bibitem [{\citenamefont {Otsu}(1979)}]{Otsu1979}%
  \BibitemOpen
\bibfield  {journal} {  }\bibfield  {author} {\bibinfo {author} {\bibfnamefont
  {N.}~\bibnamefont {Otsu}},\ }\href {\doibase 10.1109/TSMC.1979.4310076}
  {\bibfield  {journal} {\bibinfo  {journal} {IEEE Trans. Syst. Man. Cybern.}\
  }\textbf {\bibinfo {volume} {9}},\ \bibinfo {pages} {62} (\bibinfo {year}
  {1979})}\BibitemShut {NoStop}%
\bibitem [{\citenamefont {Brunel}(2000)}]{Brunel}%
  \BibitemOpen
  \bibfield  {author} {\bibinfo {author} {\bibfnamefont {N.}~\bibnamefont
  {Brunel}},\ }\href {\doibase 10.1023/A:1008925309027} {\bibfield  {journal}
  {\bibinfo  {journal} {Comput. Neurosci.}\ }\textbf {\bibinfo {volume} {8}},\
  \bibinfo {pages} {183} (\bibinfo {year} {2000})}\BibitemShut {NoStop}%
\bibitem [{Note1()}]{Note1}%
  \BibitemOpen
  \bibinfo {note} {One may alternatively collect events close to several
  references $e_{i,r\protect \tmspace +\thinmuskip {.1667em}}$,
  $e_{i,r'\protect \tmspace +\thinmuskip {.1667em}}$, etc. and concatenate
  local (and generally different) approximations of the form (\ref
  {eq:linearization_final}) yielding constraints of the same (linear) type and
  solve the regression problem via block-sparse regression \cite
  {Casadiego2017}.}\BibitemShut {Stop}%
\bibitem [{\citenamefont {van Vreeswijk}(1996)}]{Vreeswijk1996}%
  \BibitemOpen
  \bibfield  {author} {\bibinfo {author} {\bibfnamefont {C.}~\bibnamefont {van
  Vreeswijk}},\ }\href {\doibase 10.1103/PhysRevE.54.5522} {\bibfield
  {journal} {\bibinfo  {journal} {Phys. Rev. E}\ }\textbf {\bibinfo {volume}
  {54}},\ \bibinfo {pages} {5522} (\bibinfo {year} {1996})}\BibitemShut
  {NoStop}%
\bibitem [{\citenamefont {van Vreeswijk}\ and\ \citenamefont
  {Sompolinsky}(1996)}]{VanVreeswijk1996}%
  \BibitemOpen
  \bibfield  {author} {\bibinfo {author} {\bibfnamefont {C.}~\bibnamefont {van
  Vreeswijk}}\ and\ \bibinfo {author} {\bibfnamefont {H.}~\bibnamefont
  {Sompolinsky}},\ }\href
  {http://science.sciencemag.org/content/274/5293/1724.long} {\bibfield
  {journal} {\bibinfo  {journal} {Science}\ }\textbf {\bibinfo {volume} {274}}
  (\bibinfo {year} {1996})}\BibitemShut {NoStop}%
\bibitem [{\citenamefont {Timme}\ \emph {et~al.}(2002)\citenamefont {Timme},
  \citenamefont {Wolf},\ and\ \citenamefont {Geisel}}]{Timme:2002}%
  \BibitemOpen
  \bibfield  {author} {\bibinfo {author} {\bibfnamefont {M.}~\bibnamefont
  {Timme}}, \bibinfo {author} {\bibfnamefont {F.}~\bibnamefont {Wolf}}, \ and\
  \bibinfo {author} {\bibfnamefont {T.}~\bibnamefont {Geisel}},\ }\href
  {\doibase 10.1103/PhysRevLett.89.258701} {\bibfield  {journal} {\bibinfo
  {journal} {Phys. Rev. Lett.}\ }\textbf {\bibinfo {volume} {89}},\ \bibinfo
  {pages} {258701} (\bibinfo {year} {2002})}\BibitemShut {NoStop}%
\bibitem [{\citenamefont {Hansel}\ and\ \citenamefont
  {Mato}(2013)}]{Hansel2013}%
  \BibitemOpen
  \bibfield  {author} {\bibinfo {author} {\bibfnamefont {D.}~\bibnamefont
  {Hansel}}\ and\ \bibinfo {author} {\bibfnamefont {G.}~\bibnamefont {Mato}},\
  }\href {http://www.jneurosci.org/content/33/1/133.long} {\bibfield  {journal}
  {\bibinfo  {journal} {J. Neurosci.}\ }\textbf {\bibinfo {volume} {33}},\
  \bibinfo {pages} {133} (\bibinfo {year} {2013})}\BibitemShut {NoStop}%
\bibitem [{Note2()}]{Note2}%
  \BibitemOpen
  \bibinfo {note} {Reconstructions were performed on a single core of a desktop
  computer with Intel\textregistered {} Core\texttrademark {} i7-4770
  CPU@3.40GHz octocore processor and 16 GB of RAM memory . Most of execution
  time is exhausted in the computation of distances among the events.
  Reconstruction time is further determined by the speed of the numerical
  package chosen to perform least squares optimization.}\BibitemShut {Stop}%
\bibitem [{\citenamefont {English}\ \emph {et~al.}(2017)\citenamefont
  {English}, \citenamefont {McKenzie}, \citenamefont {Evans}, \citenamefont
  {Kim}, \citenamefont {Yoon},\ and\ \citenamefont
  {Buzs{\'{a}}ki}}]{English2017}%
  \BibitemOpen
  \bibfield  {author} {\bibinfo {author} {\bibfnamefont {D.~F.}\ \bibnamefont
  {English}}, \bibinfo {author} {\bibfnamefont {S.}~\bibnamefont {McKenzie}},
  \bibinfo {author} {\bibfnamefont {T.}~\bibnamefont {Evans}}, \bibinfo
  {author} {\bibfnamefont {K.}~\bibnamefont {Kim}}, \bibinfo {author}
  {\bibfnamefont {E.}~\bibnamefont {Yoon}}, \ and\ \bibinfo {author}
  {\bibfnamefont {G.}~\bibnamefont {Buzs{\'{a}}ki}},\ }\href {\doibase
  10.1016/j.neuron.2017.09.033} {\bibfield  {journal} {\bibinfo  {journal}
  {Neuron}\ }\textbf {\bibinfo {volume} {96}},\ \bibinfo {pages} {505}
  (\bibinfo {year} {2017})}\BibitemShut {NoStop}%
\bibitem [{\citenamefont {Bettstetter}\ and\ \citenamefont
  {Christian}(2002)}]{Bettstetter2002}%
  \BibitemOpen
  \bibfield  {author} {\bibinfo {author} {\bibfnamefont {C.}~\bibnamefont
  {Bettstetter}}\ and\ \bibinfo {author} {\bibnamefont {Christian}},\ }in\
  \href {\doibase 10.1145/513800.513811} {\emph {\bibinfo {booktitle} {Proc.
  3rd ACM Int. Symp. Mob. ad hoc Netw. Comput. - MobiHoc '02}}}\ (\bibinfo
  {publisher} {ACM Press},\ \bibinfo {address} {New York, New York, USA},\
  \bibinfo {year} {2002})\ p.~\bibinfo {pages} {80}\BibitemShut {NoStop}%
\bibitem [{\citenamefont {Klinglmayr}\ \emph {et~al.}(2012)\citenamefont
  {Klinglmayr}, \citenamefont {Kirst}, \citenamefont {Bettstetter},\ and\
  \citenamefont {Timme}}]{Klinglmayr2012}%
  \BibitemOpen
  \bibfield  {author} {\bibinfo {author} {\bibfnamefont {J.}~\bibnamefont
  {Klinglmayr}}, \bibinfo {author} {\bibfnamefont {C.}~\bibnamefont {Kirst}},
  \bibinfo {author} {\bibfnamefont {C.}~\bibnamefont {Bettstetter}}, \ and\
  \bibinfo {author} {\bibfnamefont {M.}~\bibnamefont {Timme}},\ }\href
  {\doibase 10.1088/1367-2630/14/7/073031} {\bibfield  {journal} {\bibinfo
  {journal} {New J. Phys.}\ }\textbf {\bibinfo {volume} {14}},\ \bibinfo
  {pages} {073031} (\bibinfo {year} {2012})}\BibitemShut {NoStop}%
\end{thebibliography}
\end{document}